\documentstyle[aps]{revtex}
\newcommand{\beg}{\begin{subequations}}
\newcommand{\eeg}{\end{subequations}}
\newcommand{\be}{\begin{equation}}
\newcommand{\ee}{\end{equation}}
\newcommand{\ba}{\begin{eqnarray}}
\newcommand{\ea}{\end{eqnarray}}
\newcommand{\bd}{\begin{displaymath}}
\newcommand{\ed}{\end{displaymath}}
\newcommand{\bi}{\begin{itemize}}
\newcommand{\ei}{\end{itemize}}
\newcommand{\ben}{\begin{enumerate}}
\newcommand{\een}{\end{enumerate}}
\newcommand{\bed}{\begin{description}}
\newcommand{\eed}{\end{description}}
\newcommand{\ie}{{\it i.e. }}
\newcommand{\eg}{{\it e.g. }}
\newcommand{\et}{{\it et al.}}

\newcommand{\br}[1]{(\ref{#1})}

\newcommand{\Hm}{{\cal H}}

\newcommand{\up}{\uparrow}
\newcommand{\dn}{\downarrow}

\DeclareSymbolFont{boldletters}     {OML}{cmm} {b}{it}
\DeclareSymbolFont{boldoperators}   {OT1}{cmr}{b}{n}
\DeclareSymbolFontAlphabet{\mathbold}{boldletters}
\DeclareMathSymbol{\bfalpha}{\mathord}{boldletters}{"0B}
\DeclareMathSymbol{\bfbeta}{\mathord}{boldletters}{"0C}
\DeclareMathSymbol{\bfgamma}{\mathord}{boldletters}{"0D}
\DeclareMathSymbol{\bfdelta}{\mathord}{boldletters}{"0E}
\DeclareMathSymbol{\bfepsilon}{\mathord}{boldletters}{"0F}
\DeclareMathSymbol{\bfzeta}{\mathord}{boldletters}{"10}
\DeclareMathSymbol{\bfeta}{\mathord}{boldletters}{"11}
\DeclareMathSymbol{\bftheta}{\mathord}{boldletters}{"12}
\DeclareMathSymbol{\bfiota}{\mathord}{boldletters}{"13}
\DeclareMathSymbol{\bfkappa}{\mathord}{boldletters}{"14}
\DeclareMathSymbol{\bflambda}{\mathord}{boldletters}{"15}
\DeclareMathSymbol{\bfmu}{\mathord}{boldletters}{"16}
\DeclareMathSymbol{\bfnu}{\mathord}{boldletters}{"17}
\DeclareMathSymbol{\bfxi}{\mathord}{boldletters}{"18}
\DeclareMathSymbol{\bfpi}{\mathord}{boldletters}{"19}
\DeclareMathSymbol{\bfrho}{\mathord}{boldletters}{"1A}
\DeclareMathSymbol{\bfsigma}{\mathord}{boldletters}{"1B}
\DeclareMathSymbol{\bftau}{\mathord}{boldletters}{"1C}
\DeclareMathSymbol{\bfupsilon}{\mathord}{boldletters}{"1D}
\DeclareMathSymbol{\bfphi}{\mathord}{boldletters}{"1E}
\DeclareMathSymbol{\bfchi}{\mathord}{boldletters}{"1F}
\DeclareMathSymbol{\bfpsi}{\mathord}{boldletters}{"20}
\DeclareMathSymbol{\bfomega}{\mathord}{boldletters}{"21}
\DeclareMathSymbol{\bfvarepsilon}{\mathord}{boldletters}{"22}
\DeclareMathSymbol{\bfvartheta}{\mathord}{boldletters}{"23}
\DeclareMathSymbol{\bfvarpi}{\mathord}{boldletters}{"24}
\DeclareMathSymbol{\bfvarrho}{\mathord}{boldletters}{"25}
\DeclareMathSymbol{\bfvarsigma}{\mathord}{boldletters}{"26}
\DeclareMathSymbol{\bfvarphi}{\mathord}{boldletters}{"27}
\DeclareMathSymbol{\bfGamma}{\mathalpha}{boldoperators}{"00}
\DeclareMathSymbol{\bfDelta}{\mathalpha}{boldoperators}{"01}
\DeclareMathSymbol{\bfTheta}{\mathalpha}{boldoperators}{"02}
\DeclareMathSymbol{\bfLambda}{\mathalpha}{boldoperators}{"03}
\DeclareMathSymbol{\bfXi}{\mathalpha}{boldoperators}{"04}
\DeclareMathSymbol{\bfPi}{\mathalpha}{boldoperators}{"05}
\DeclareMathSymbol{\bfSigma}{\mathalpha}{boldoperators}{"06}
\DeclareMathSymbol{\bfUpsilon}{\mathalpha}{boldoperators}{"07}
\DeclareMathSymbol{\bfPhi}{\mathalpha}{boldoperators}{"08}
\DeclareMathSymbol{\bfPsi}{\mathalpha}{boldoperators}{"09}
\DeclareMathSymbol{\bfOmega}{\mathalpha}{boldoperators}{"0A}

\DeclareMathSymbol{\bfslA}{\mathalpha}{boldletters}{"41}
\DeclareMathSymbol{\bfslB}{\mathalpha}{boldletters}{"42}
\DeclareMathSymbol{\bfslC}{\mathalpha}{boldletters}{"43}
\DeclareMathSymbol{\bfslD}{\mathalpha}{boldletters}{"44}
\DeclareMathSymbol{\bfslE}{\mathalpha}{boldletters}{"45}
\DeclareMathSymbol{\bfslF}{\mathalpha}{boldletters}{"46}
\DeclareMathSymbol{\bfslG}{\mathalpha}{boldletters}{"47}
\DeclareMathSymbol{\bfslH}{\mathalpha}{boldletters}{"48}
\DeclareMathSymbol{\bfslI}{\mathalpha}{boldletters}{"49}
\DeclareMathSymbol{\bfslJ}{\mathalpha}{boldletters}{"4A}
\DeclareMathSymbol{\bfslK}{\mathalpha}{boldletters}{"4B}
\DeclareMathSymbol{\bfslL}{\mathalpha}{boldletters}{"4C}
\DeclareMathSymbol{\bfslM}{\mathalpha}{boldletters}{"4D}
\DeclareMathSymbol{\bfslN}{\mathalpha}{boldletters}{"4E}
\DeclareMathSymbol{\bfslO}{\mathalpha}{boldletters}{"4F}
\DeclareMathSymbol{\bfslP}{\mathalpha}{boldletters}{"50}
\DeclareMathSymbol{\bfslQ}{\mathalpha}{boldletters}{"51}
\DeclareMathSymbol{\bfslR}{\mathalpha}{boldletters}{"52}
\DeclareMathSymbol{\bfslS}{\mathalpha}{boldletters}{"53}
\DeclareMathSymbol{\bfslT}{\mathalpha}{boldletters}{"54}
\DeclareMathSymbol{\bfslU}{\mathalpha}{boldletters}{"55}
\DeclareMathSymbol{\bfslV}{\mathalpha}{boldletters}{"56}
\DeclareMathSymbol{\bfslW}{\mathalpha}{boldletters}{"57}
\DeclareMathSymbol{\bfslX}{\mathalpha}{boldletters}{"58}
\DeclareMathSymbol{\bfslY}{\mathalpha}{boldletters}{"59}
\DeclareMathSymbol{\bfslZ}{\mathalpha}{boldletters}{"5A}

\DeclareMathSymbol{\bfsla}{\mathalpha}{boldletters}{"61}
\DeclareMathSymbol{\bfslb}{\mathalpha}{boldletters}{"62}
\DeclareMathSymbol{\bfslc}{\mathalpha}{boldletters}{"63}
\DeclareMathSymbol{\bfsld}{\mathalpha}{boldletters}{"64}
\DeclareMathSymbol{\bfsle}{\mathalpha}{boldletters}{"65}
\DeclareMathSymbol{\bfslf}{\mathalpha}{boldletters}{"66}
\DeclareMathSymbol{\bfslg}{\mathalpha}{boldletters}{"67}
\DeclareMathSymbol{\bfslh}{\mathalpha}{boldletters}{"68}
\DeclareMathSymbol{\bfsli}{\mathalpha}{boldletters}{"69}
\DeclareMathSymbol{\bfslj}{\mathalpha}{boldletters}{"6A}
\DeclareMathSymbol{\bfslk}{\mathalpha}{boldletters}{"6B}
\DeclareMathSymbol{\bfsll}{\mathalpha}{boldletters}{"6C}
\DeclareMathSymbol{\bfslm}{\mathalpha}{boldletters}{"6D}
\DeclareMathSymbol{\bfsln}{\mathalpha}{boldletters}{"6E}
\DeclareMathSymbol{\bfslo}{\mathalpha}{boldletters}{"6F}
\DeclareMathSymbol{\bfslp}{\mathalpha}{boldletters}{"70}
\DeclareMathSymbol{\bfslq}{\mathalpha}{boldletters}{"71}
\DeclareMathSymbol{\bfslr}{\mathalpha}{boldletters}{"72}
\DeclareMathSymbol{\bfsls}{\mathalpha}{boldletters}{"73}
\DeclareMathSymbol{\bfslt}{\mathalpha}{boldletters}{"74}
\DeclareMathSymbol{\bfslu}{\mathalpha}{boldletters}{"75}
\DeclareMathSymbol{\bfslv}{\mathalpha}{boldletters}{"76}
\DeclareMathSymbol{\bfslw}{\mathalpha}{boldletters}{"77}
\DeclareMathSymbol{\bfslx}{\mathalpha}{boldletters}{"78}
\DeclareMathSymbol{\bfsly}{\mathalpha}{boldletters}{"79}
\DeclareMathSymbol{\bfslz}{\mathalpha}{boldletters}{"7A}

\DeclareMathSymbol{\bfell}{\mathalpha}{boldletters}{"60}

\begin{document}
\draft

\title{Quantum Transport in Disorderd Mesoscopic Ferromagnetic Films}
\author{Philip A.E. Jonkers\cite{email}, Steven J. Pickering and Hans De
Raedt\cite{email2}}
\address{Rijksuniversiteit Groningen, 9747 AG Groningen, The Netherlands}
\author{Gen Tatara\cite{email3}}
\address{
Max Planck Institut fur Mikrostrukturphysik
Weinberg 2, 06120 Halle, Germany
\\
and\\
Graduate School of Science, Osaka University, Toyonaka, Osaka
560-0043, Japan}
\maketitle
\begin{abstract}%
The effect of impurity and domain-wall scattering on the electrical
conductivity of disordered mesoscopic
magnetic thin films is studied by use of computer simulation.
The results indicate a reduction of resistivity due to a domain wall,
which is consistent with the explanation in terms of the dephasing caused
by domain wall.
\end{abstract}

The electrical transport properties of ferromagnetic metals have attracted
much interest
recently see \eg \cite{TODOROV}-\cite{GREGG}. In the present work we study
the quantum transport in mesoscopic wires that contain
a magnetic domain wall. The motion of the electrons passing through a wire
that
contain a magnetic domain wall is affected by various physical processes.
As the electron
approaches the domain wall it experiences a change in potential energy,
leading to a reflection
and hence to a reduction of the conductivity. However, unless the domain
wall is unrealistically
narrow (compared to the Fermi wavelength of the electrons) this reduction
has been shown to be
negligibly small \cite{CABRERA} in the case of a spin-independent collision
time. In the presence of a domain wall the spin of the electron will
change as the electron passes through the wire. This rotation will lead to
a mixing of spin-up
and spin-down components. Assuming that the (Boltzmann) collision time
is spin-dependent, this mixing then results in an increase of the
resistivity, a scenario that
has been proposed \cite{LEVY} to explain the experimental results on thin
Co films at room
temperature \cite{GREGG}. Spin dependent scattering is the essential
ingredient in models for electron
transport in magnetic materials that exhibit giant magnetoresistance (GMR)
\cite{BAIBICH}-\cite{PARKIN}.

In disordered systems at low temperatures the quantum interference, which
becomes important as a result of random spin-{\it in}dependent impurity
scattering, also influences strongly the
electron transport properties. Theoretical work \cite{TATARA} has shown
that the domain wall
suppresses the interference (and thus weak localization) due to impurity
scattering,
resulting in
a decrease of the resistivity. Very recently there have been several
experimental studies of a resistivity in
a mesoscopic wire of ferromagnetic metals \cite{RUEDIGER}-\cite{OTANI}. The
results
suggest a reduction of resistivity due to a domain wall, and interestingly
the effect increases by lowering the temperature; below 50 K
\cite{RUEDIGER}, and 20 K \cite{OTANI} respectively. This reduction might
be related to the quantum decoherence
caused by the wall. But other classical mechanisms of the reduction have also
been proposed as well \cite{RUEDIGER} and further studies are needed to
clarify its origin. The purpose of the present paper is to study the
interplay of the domain wall and
spin-independent impurity
scattering in more detail and to compare quantitatively the theoretical
prediction of the
Kubo-formula approach with first-principle quantum mechanical calculations.

The geometry of the model system is shown in Fig.\ref{system}. The electrons
are assumed to move in a two-dimensional metallic strip with a single
magnetic domain wall. The Hamiltonian for this model reads
\be \label{Schr}
 \Hm = \frac{1}{2m^\ast}\Bigl({\bf p} - e {\bf A}/c\Bigr)^2 - \mu_B
\bfsigma \cdot {\bf M} + V,
\ee
where ${\bf p} = (p_x,p_y)$ is the momentum operator of the electron with
effective mass $m^\ast$,
$\bfsigma = (\sigma^x,\sigma^y,\sigma^z)$ denote the Pauli spin matrices.
${\bf M} = {\bf M}(x,y)$
describes the magnetization in the material and $V=V(x,y)$ represents the
potential due
to non-magnetic impurities. We neglect the vector potential ${\bf A}$
resulting from the
sum of the atomic magnetic-dipole contributions because in the case of a
thin wire, it has little effect on the electron transport.

Following \cite{CABRERA},\cite{TATARA} we assume that the magnetic domain
wall can be described by
\be
 M_x(x,y) = M_0 {\rm sech}\Bigl( \frac{x-x_0}{\lambda_w} \Bigr)
\ee
and
\be
 M_z(x,y) = M_0 {\rm tanh}\Bigl( \frac{x-x_0}{\lambda_w} \Bigr),
\ee
with $x_0$ the center of the domain wall and $\lambda_w$ measures its extent.
Note that $M_z^2(x,y) + M_x^2(x,y) = M_0^2$ so that at each point $(x,y)$
the magnetization is a
constant. For a schematic picture of how the magnetization changes with $x$
see Fig.\ref{system}.

As a model for each impurity we take a square potential barrier, \ie
\be
 V_n(x,y) = \left\{ \begin{array}{cr} 0~,\quad & (x,y) \not\in S_n \\ V_0~,
\quad & (x,y) \in S_n \\ \end{array} \right.
\ee
where $S_n$ denote the area of square with label $n$. The position of the
square is drawn from a
uniform random distribution, rescaled to an area of size  $L_x \times L_y$
(see Fig.\ref{system}).
The concentration of impurities, $c$ is given by $c = \sum_{n=1}^N S_n/(L_x
L_y)$ where $N$ denotes
the total number of impurities. The potential entering in Eq. \br{Schr} is
given by $V = V(x,y) = \sum_{n=1}^N V_n(x,y)$.

We will follow two routes to study the effect of the domain wall on the
electrical conductivity:
1) By solving the time-dependent Schr\"odinger equation (TDSE) and
2) through an extension of the Kubo-formula-based theory of Tatara and
Fukuyama \cite{TATARA}.
The results of these two fundamentally different approaches can be compared
by making use of the
Landauer formula \cite{LANDAUER},\cite{DATTA}
relating the conductivity $\sigma$ to the tranmission coefficient $T$.

In the TDSE approach the procedure to calculate the transmission
coefficient $T$ consists of three
steps. First the incoming electrons are represented by a wave packet with
average momentum
$\langle {\bf p} \rangle = \hbar \bfslk = (\hbar k_F,0)$. For concreteness
we take this intitial state to represent electrons with spin up only, \ie
\be
 \Psi(x,y,t=0) = (\psi_\up(x,y,t=0),\psi_\dn(x,y,t=0)) =
(\psi_\up(x,y,t=0),0),
\ee
and $\int dx dy |\Psi(x,y,t=0)|^2=1$.
The second step involves the solution of the TDSE
\be \label{TdSe}
 i \hbar \frac{\partial \Psi(x,y,t)}{\partial t} = \Hm \Psi(x,y,t)
\ee
for sufficiently long times. The method we use to solve the TDSE has been
described at length
elsewhere \cite{DERAEDT2},\cite{DERAEDT1}, so we omit details here.  As
indicated in Fig.\ref{system}, we place
imaginary detection screens at various $x-$positions. The purpose of each
screen is to record
the accumulated current that passes through it (the wave function is not
modified by this detection
process). Dividing the transmitted current (detector 2, see
Fig.\ref{system}) by the incident
current (detector 1) yields the transmission coefficient $T$.
As the simulation package \cite{DERAEDT2},\cite{DERAEDT1} that we use
solves the TDSE subjected to Dirichlet boundary
conditions, some precautions have to be taken in order to suppress
artifacts due to reflections
from the boundaries at $x=0,x=L$. We have chosen to add to $V$, an
imaginary linear potential
that is non-zero near the edges of the sample, as indicated by the gray
strips in Fig.\ref{system},
and found that the absorption of intensity that results is adequate for the
present purpose.

For numerical work it is convenient to rewrite the TDSE \br{TdSe} in a
dimensionless form.
Taking the Fermi wavelength $\lambda_F$ as the characteristic length scale
of the electrons, the energy is measured
in units of the Fermi-energy $E_F = h^2 / (2 m \lambda_F^2)$ and time in
units of $\hbar/E_F$.
For our model simulations we have taken $L=100~\lambda_F$, $L_y =
6.5~\lambda_F$, $\mu_B M_0 =
0.4~E_F$, $V_0 = 100~E_F$ and $S_n = 0.25~\lambda_F^2$.

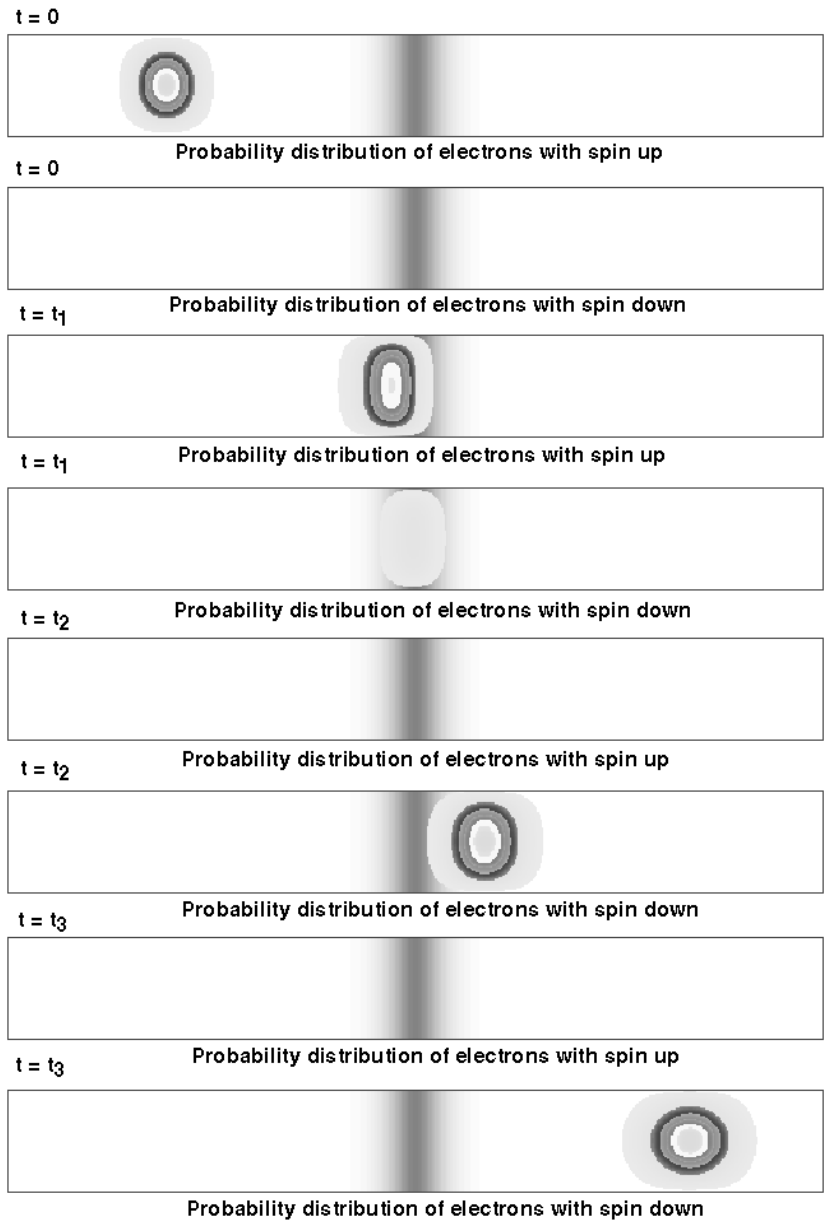
\begin{figure}[htbp]
 \begin{center}
  \caption{Snapshots of the time evolution of the electron wave packet moving
 through an impurity free mesoscopic wire containing a domain wall with
$\lambda_w = 2 \lambda_F$ (represented by the smooth gray area),
taken at $t_1=75~\hbar/E_F$, $t_2 = 100~\hbar/E_F$ and $t_3 =
150~\hbar/E_F$.} \label{snap1}
 \end{center}
\end{figure}

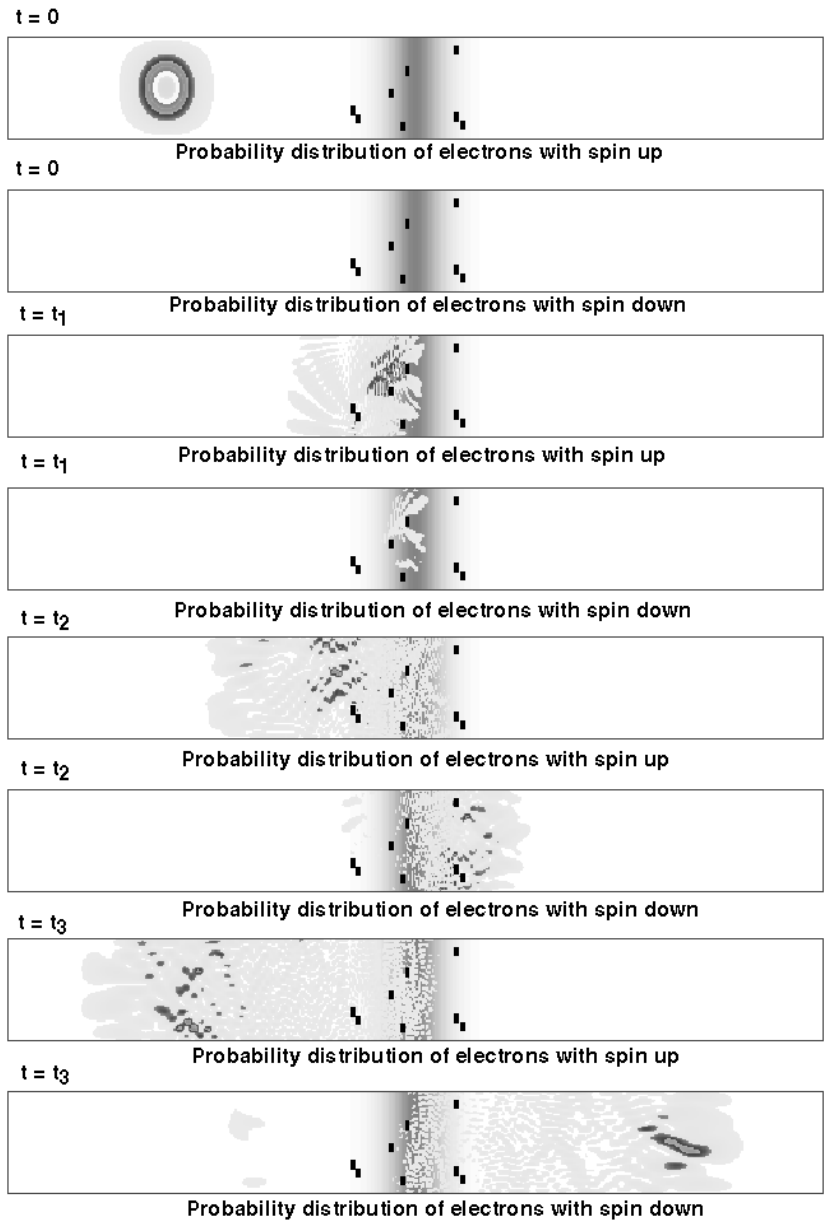
\begin{figure}[htbp]
 \begin{center}
  \caption{Snapshots of the time evolution of the electron wave packet
moving through a mesoscopic wire with impurities (represented by small
black dots)
with an impurity concentration $c=2 \%$ containing a domain wall
($\lambda_w = 2 \lambda_F$), taken at $t_1=75~\hbar/E_F$, $t_2 =
100~\hbar/E_F$ and $t_3 = 150~\hbar/E_F$.} \label{snap2}
 \end{center}
\end{figure}

In Figs.\ref{snap1} and \ref{snap2} we show some snapshots of the
probability distribution for the spin-up (top)
and spin-down (bottom) part of the electron wave, moving through an
impurity-free region.
Initially at $t=0$, the probability for having electrons with
spin-down is zero.
As the wave moves to the right, the $M_x$ component of the magnetization
causes the spin to rotate,
resulting in a conversion of electrons with spin-up into electrons with
spin-down.
For realistic values of the strength (\ie $\mu_B M_0 < E_F$) and width of
the domain wall (\ie $\lambda_w > \lambda_F$) the conversion will be almost
100 $\%$ (for all practical purposes), which leads to a negligibly small
reflection \cite{CABRERA}. We have chosen $\lambda_w =
2~\lambda_F,\ldots,16~\lambda_F$,
which may be reasonable in the case of a very narrow wire or a strong
anisotropy.

In the presence of impurities two new effects appear:
\begin{enumerate}
 \item As a result of the scattering by the potential barriers electrons
will be reflected, leading
to a reduction of the transmission coefficient in the sense of Boltzmann
transport. At the same time interference among scattered electrons leads
to weak localization, and this quantum mechanical effect also suppresses
the transmission. Obviously these effects are present in the absence
of a domain wall as well.
 \vspace{-0.3 cm}
 \item As a result of the presence of the domain wall, electrons that are
backscattered {\it and} have
their spin reversed due to the wall, no longer interfere with electrons whose
spin is unchanged.
Hence the effect of the domain wall is to reduce the enhanced
backscattering due to the interference.
On the basis of this argument it is to be expected that in the presence of
a domain wall the
transmission coefficient can be larger than in the absence of it.
\end{enumerate}

In our simulations the contribution due to quantum interference effects
resulting from the presence
of the domain wall can be separated from all other contributions by a
simple procedure: We
compute the ratio of the transmission with $(T)$ and without $(T_0)$ a domain
wall.

Some representative results of our calculations are depicted in
Figs.\ref{figt0}-\ref{Walls2}.
The simulation data shown are obtained from a single realization of the
impurity distribution.
No ensemble averaging of the transmission coefficient has been performed.
The transmission in the absence of the wall ($T_0$) is plotted in Fig.
\ref{figt0}
as a function of impurity concentration in the case of $L_x=16$.
In Figs. \ref{Imps1} and \ref{Imps2} we show the ratio $T/T_0$ as a function
of the impurity
concentration $c$, for $L_x=8$ and $L_x=16$ respectively. The two sets of
simulation data in Fig. \ref{Imps1} correspond to different
impurity configurations,
and the difference between the two is due to a different interference pattern.
The enhancement alluded to above is
clearly present. The effect of conversion of the electron spin by the wall
is amplified considerably by the quantum interference at larger impurity
concentration.
The larger the scattering the more effective the domain wall is in converting
electrons with spin-up into electrons with spin-down.

In Figs.\ref{Walls1} and \ref{Walls2} we present results for domain walls
of different width
$\lambda_w$, keeping fixed the area in which the impurities are present
($L_x=4$, and $L_x=8$
respectively). The net result of increasing $\lambda_w$ in this case is to
reduce the
effectiveness of the $M_x \sigma^x$ term in the Hamiltonian. Indeed by
increasing $\lambda_w$,
$M_x(x,y)$ becomes more smooth, hence less effective in the sense that less
electrons flip their spin.

Let us compare these results with the analytical result based on Kubo
formula, which is obtained by extending the theory of Tatara and Fukuyama
\cite{TATARA}.
In the absence of domain wall the conductivity in two dimensions with
the effect of weak localization taken into account is given by
\begin{eqnarray}
	\sigma_0 &=& \frac{e^2 n\tau}{m}
- \frac{2e^2}{\pi\hbar} \frac{1}{V}\sum_q	\frac{1}{q^2}
\nonumber\\
&=& \frac{e^2}{h}n\lambda_F l
	\left(1-\frac{\lambda_F}{l}\frac{2}{\pi^3}\frac{L_x}{L_y}
    \right),
	\label{sigma0}
\end{eqnarray}
where $n$ is the electron density, $\tau$ and $l\equiv (\hbar
k_F\tau/m)$ being the elastic lifetime and the mean free path,
respectively.
We have carried out the $q$-summation in one dimension, since $L_y$ is
much smaller than the inelastic diffusion length in the absence of the
wall, which should be regarded as infinity in the simulation here.
The transmission coefficient $T_0$ is related to the conductivity by
$\sigma_0=(e^2/h)(L_x/L_y)[T_0/(1-T_0)]$ and thus
\begin{equation}
	T_0 \simeq \frac{\beta}{\beta+\nu c} \left[1-\frac{\nu
c^2}{\beta+\nu c}
	\frac{2}{\pi^3}\frac{1}{\alpha} \frac{L_x}{L_y} \right],
	\label{T0def}
\end{equation}
where $\beta\equiv n \lambda_F^2 \alpha$,  $\nu\equiv(L_x/L_y)$ and the mean
free
path is related to $c$ through $l\equiv \alpha\lambda_F /c$.
We treat $\alpha$ and $\beta$ as fitting parameters.
The solid curve in Fig. \ref{figt0} is obtained for $\alpha=0.05$ and
$\beta=6$ (or equivalently $l\sim 0.5\lambda_F\simeq 3 k_F^{-1}$ for
$c=0.1~\%$,
which appears to be reasonable).
The dotted line is the classical contribution to $T_0$ (\ie the first term in
\br{T0def}) and it is larger than $T_0$ at large $c$.

In the presence of a domain wall the conductivity is expressed as
\begin{equation}
 \sigma= \frac{e^2}{h}nl\lambda_F
	\left[1-\frac{1}{2\pi^2}\frac{\lambda_F^2}{\lambda_w L}
	-\frac{2}{\pi^2}\frac{\lambda_F}{l}
	\left(\frac{L_w}{L_y}\tan^{-1}\frac{L_x}{\pi L_w}\right) \right],
	\label{sigma}
\end{equation}
where the second term is the classical contribution from the wall
reflection and the third term is a weak localization correction with
the effect of the wall included.
The effect of the wall is to cause dephasing among the electron
as is represented by
the inelastic diffusion length, $L_w\equiv
\sqrt{D\tau_w}$.
Here $\tau_w$ is the inelastic lifetime due to the spin-flip
scattering by the wall,
$\tau_w^{-1} \equiv (\lambda_F E_F)^2 / (24 \pi^2 \lambda_w L_x \Delta^2
\tau)$ ($\Delta \equiv \mu_B M_0$ denoting the Zeeman
splitting)\cite{TATARA}.
The expression of $T/T_0$ is obtained as
\begin{equation}
	\frac{T}{T_0}=1+\frac{\nu c^2}{\beta+\nu c} \frac{1}{\alpha}
\left[ \frac{2}{\pi^3}\frac{L_x}{L_y}\left(1-\frac{\pi L_w}{L_x}
\tan^{-1}\frac{L_x}{\pi L_w} \right)
-\frac{1}{2\pi^2}\frac{\lambda_F^2}{\lambda_w L_x}  \right].
	\label{tt0}
\end{equation}
The result is plotted as solid lines in Figs. \ref{Imps1}-\ref{Walls2}.
The classical contribution (the last term) is negligibly small
compared with the quantum correction in the region we are interested,
and thus the enhancement of the transmission by the wall is seen.
We have used the same value of parameter $\beta=6$, but with
different $\alpha$ ($\alpha=0.05 $ for  Fig. \ref{Imps1} but $\alpha=0.02$
for Figs. \ref{Imps2}-\ref{Walls2}). We think this dependence of
$\alpha$ on $L_x$ is due to the ambiguity in
relating the mean free path in Kubo formula to $c$ in the simulation.
Results of eq. (\ref{tt0}) thus obtained explain the simulation data
well.

{\bf Acknowledgements} \\
This work was partially supported by the ``Stichting Nationale Computer
Faciliteiten (NCF)'',
the NWO Priority Program on Massive Parallel Processing and a Grant-in-Aid
for Scientific
Research of the Japanese Ministry of Education, Science and Culture.

\begin{figure}[htbp]
 \begin{center}
  \caption{The geometry of the simulation model of a mesoscopic metallic wire
containing a magnetic domain wall of width $\lambda_w$. Black squares:
Impurities distributed randomly over an area of size $L_x \times L_y$. The
gray stripes
at the edges indicate regions where electrons entering these regions are
being absorbed. The detector screens 1 and 2 measure the electrical current
through these
screens. Also shown is a schematic diagram of the magnetization inside the
strip.} \label{system}
 \end{center}
\end{figure}

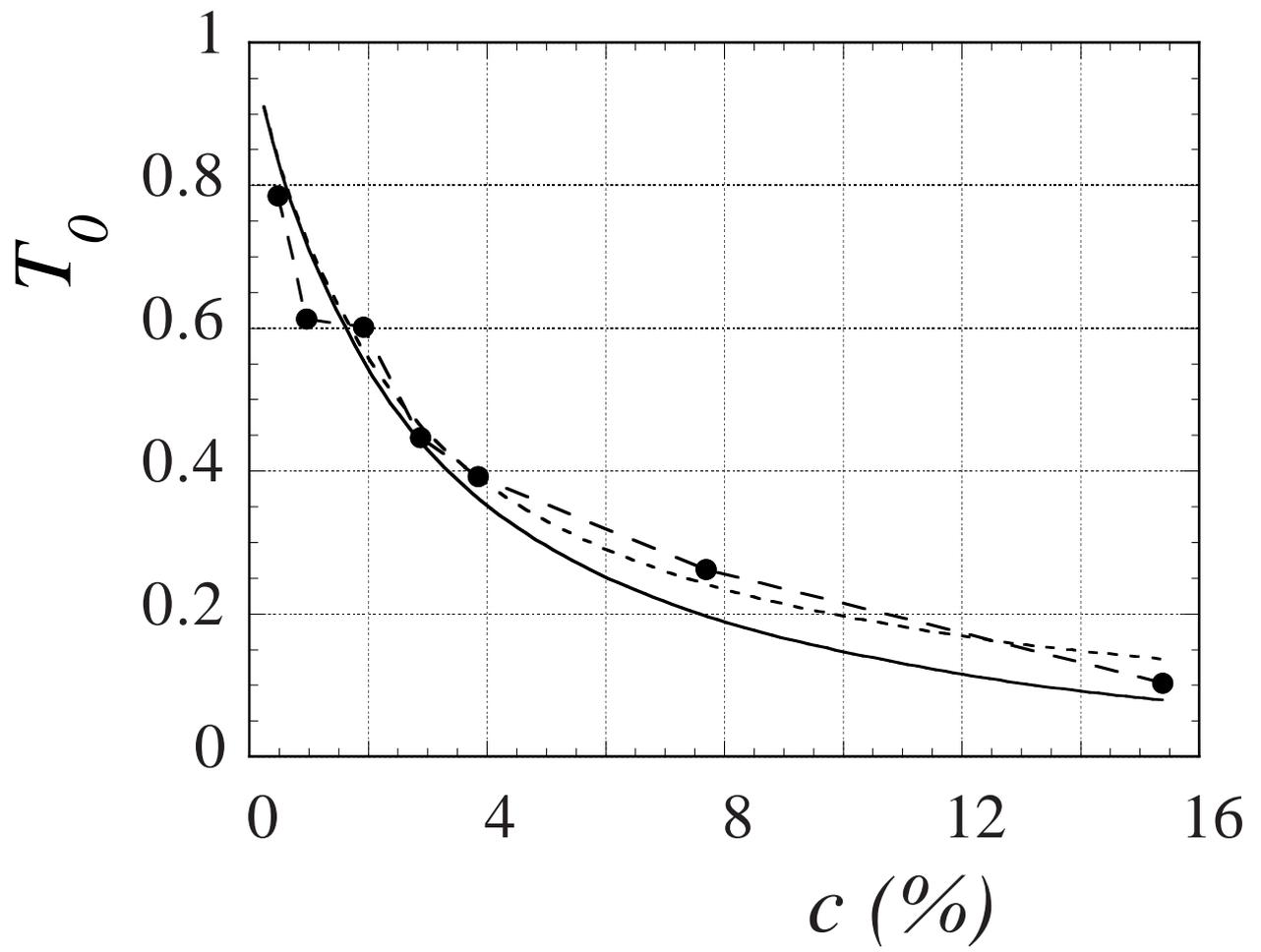
\begin{figure}[tbp]
 \begin{center}
  \caption{ Transmission in the absence of domain wall, $T_0$,
  as a function of impurity concentration
$c$ for the case of
$L_x=8~\lambda_F$. Solid and dotted line denotes the result of Kubo
formula with and without the weak localization correction taken into
account, respectively. The effect of weak localization lowers the
transmission at large $c$.
Parameters are $\alpha=0.05$ and $\beta=6$ (see \br{T0def}).
} \label{figt0}
 \end{center}
\end{figure}
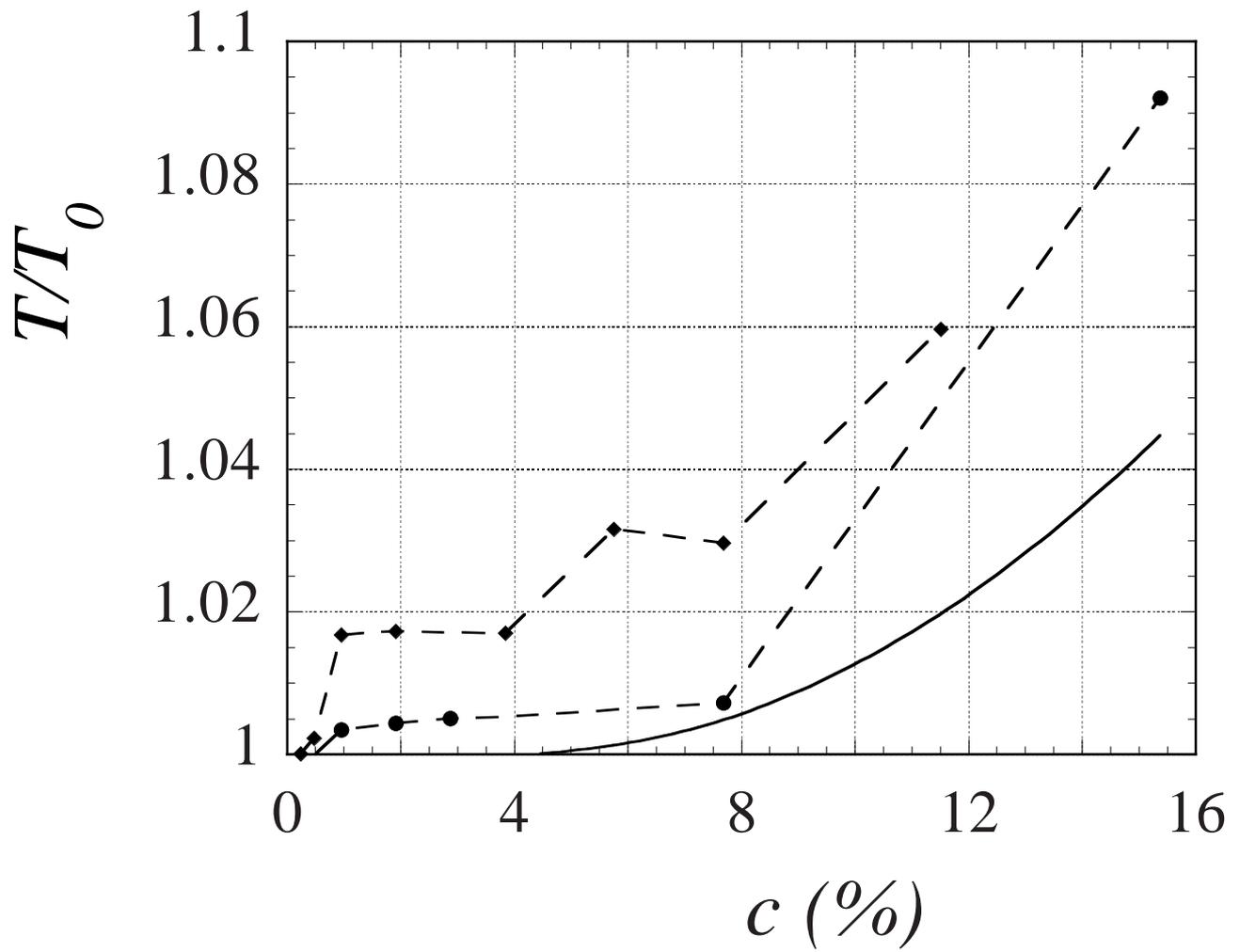
\begin{figure}[tbp]
\begin{center}
  \caption{Relative enhancement $T/T_0$ of the transmission resulting from
the presence of the domain wall as a function of impurity concentration
$c$. The width of the domain wall is $\lambda_w=2~\lambda_F$ and
$L_x=16~\lambda_F$ (see Fig.\ref{system}). Simulation data for different
impurity configurations
are represented by diamonds and circles (the dashed and dotted line are
guides to the eye only). Also shown is the theoretical result \br{tt0} with
 $\alpha=0.05$ and $\beta=6$. (solid line).} \label{Imps1}
\end{center}
\end{figure}
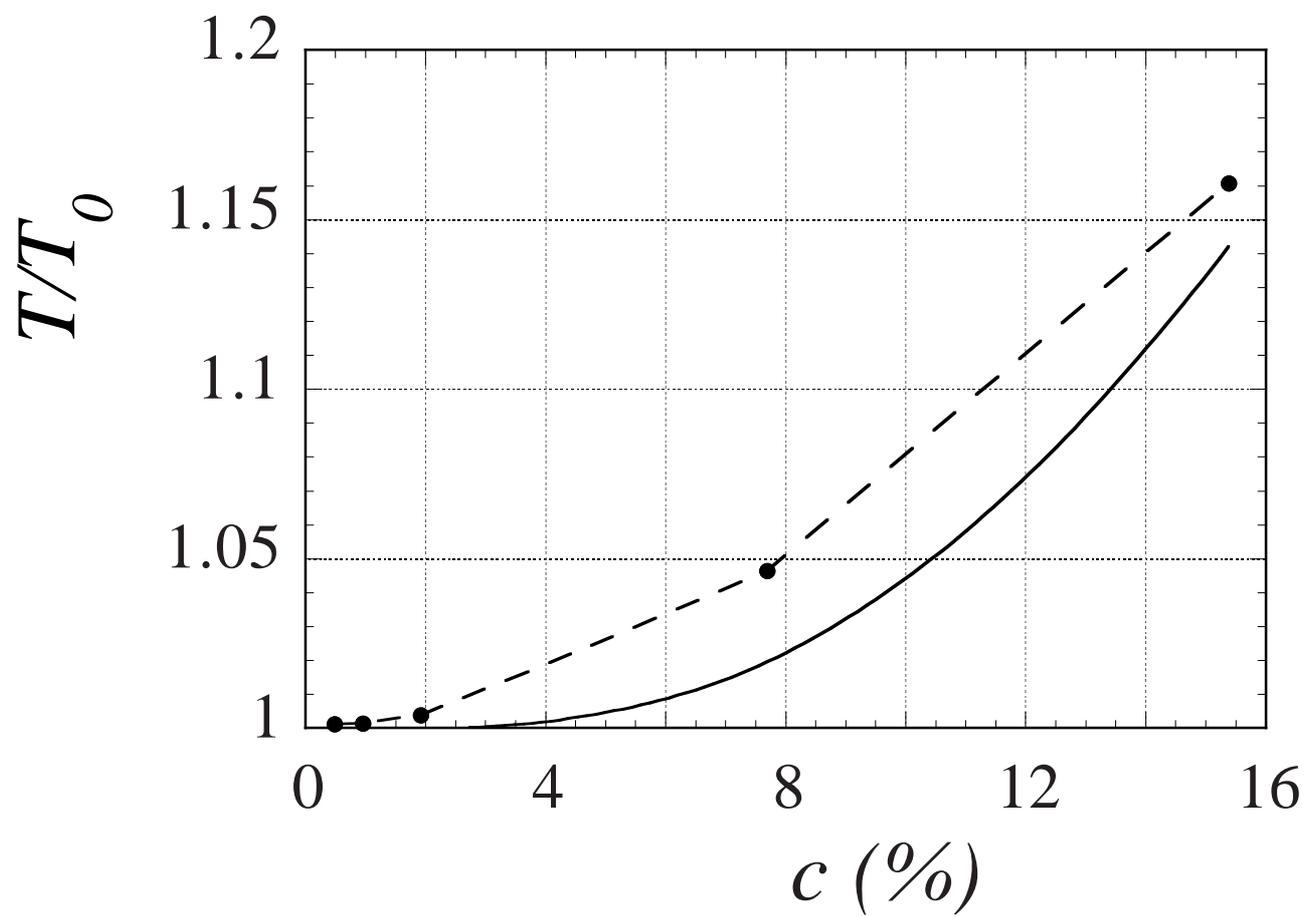
\begin{figure}[tbp]
 \begin{center}
 \vspace{0.5 cm}
  \caption{Relative enhancement $T/T_0$ of the transmission resulting from
the presence of the domain wall as a function of impurity concentration $c$.
The width of the domain wall $\lambda_w=2~\lambda_F$ and $L_x=8~\lambda_F$
(see Fig.\ref{system}). Circles: simulation data; solid line: theoretical
result \br{tt0} ($\alpha = 0.02$, $\beta=6$).} \label{Imps2}
 \end{center}
\end{figure}

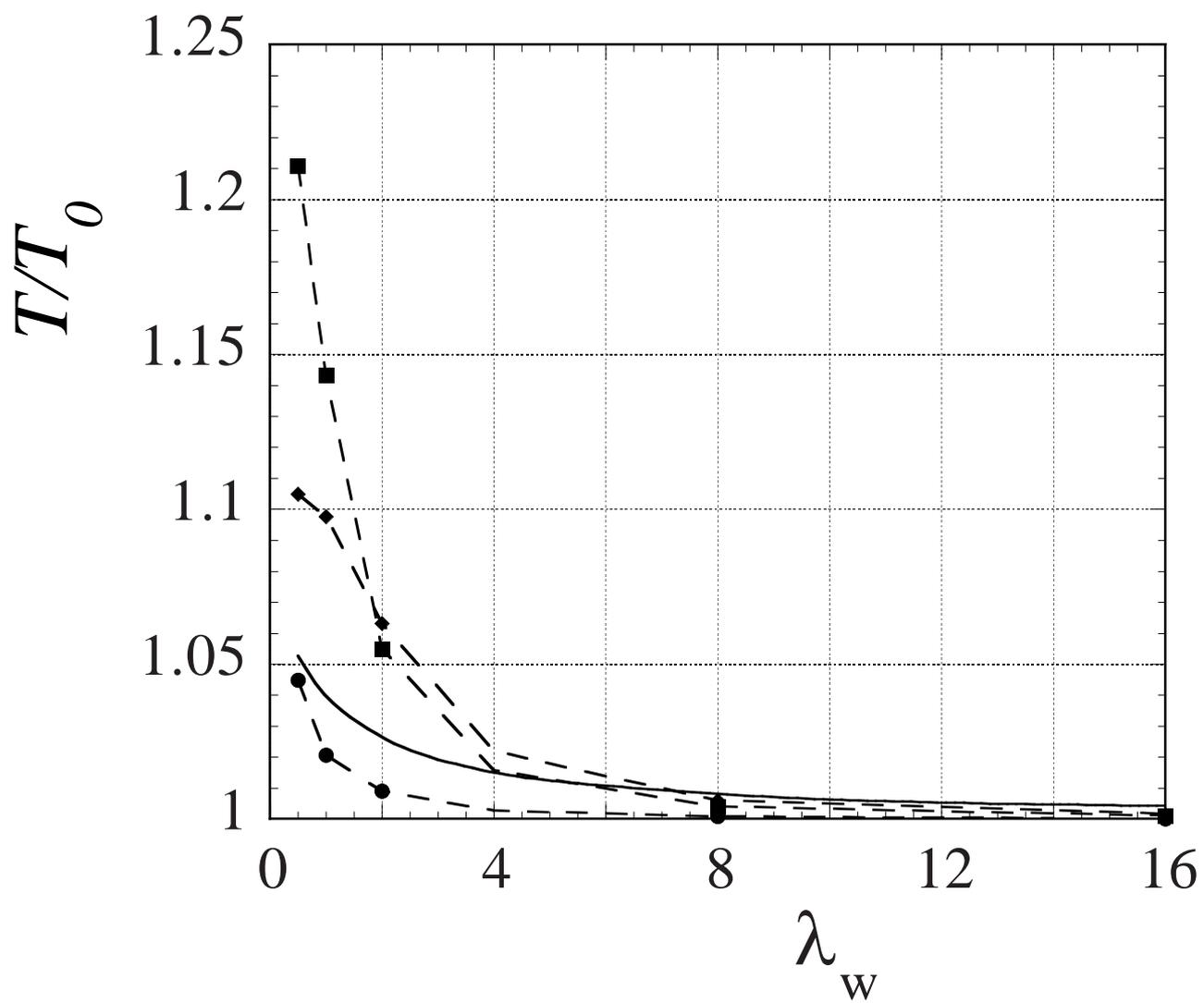
\begin{figure}[htbp]
\begin{center}
  \caption{Relative enhancement $T/T_0$ of the transmission as a function of
 the width $\lambda_w$ of the domain wall for various impurity concentrations
 $c$ and $L_x=4~\lambda_F$. The circles, squares and diamonds correspond to
$c = 3.85~\%$, $c = 7.69~\%$ and $c=15.38~\%$ respectively.
The solid line depicts the theoretical result for $c = 15.38~\%$
($\alpha = 0.02$, $\beta=6$).} \label{Walls1}
\end{center}
\end{figure}
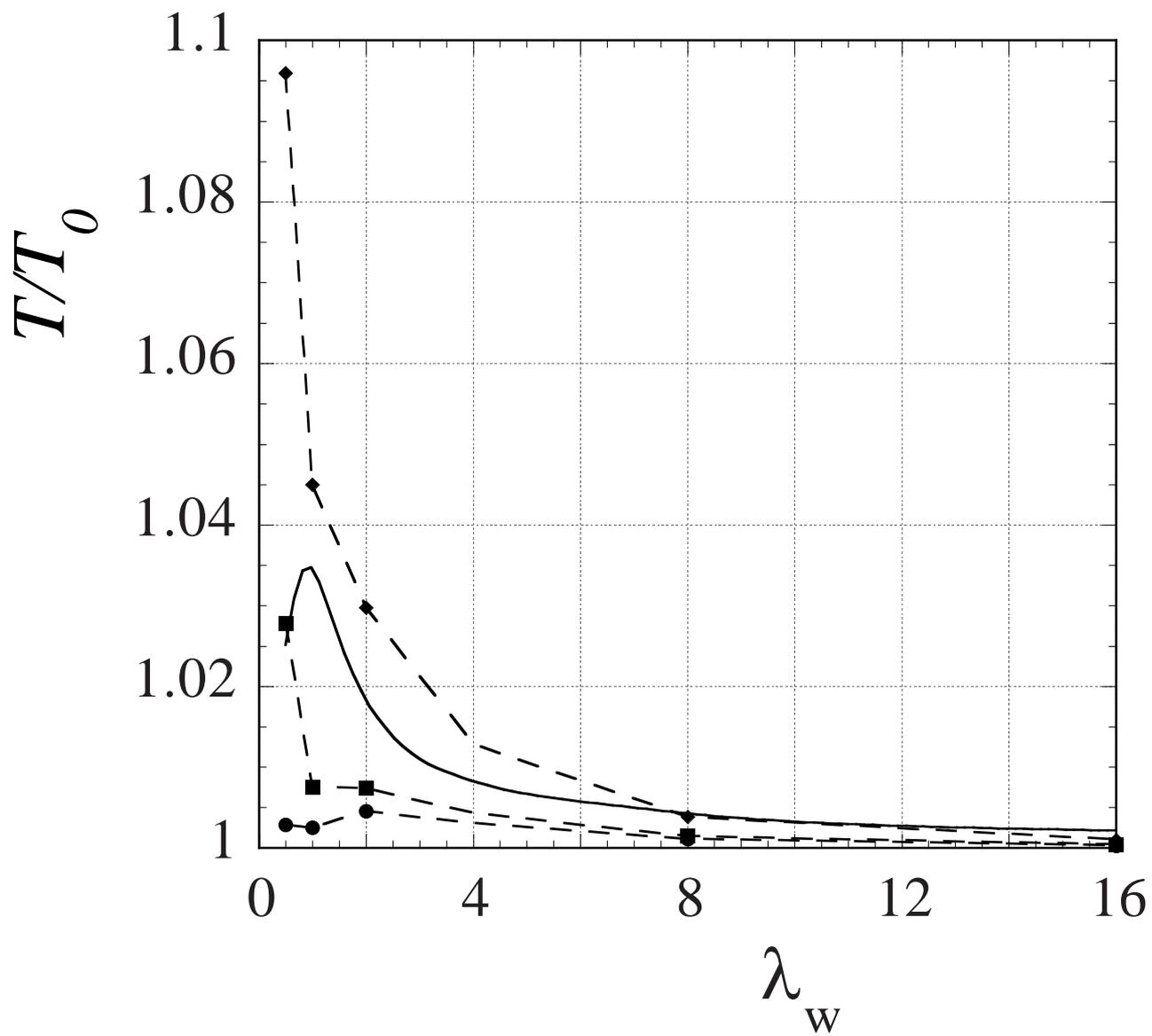
\begin{figure}[htbp]
 \begin{center}
 \vspace{0.5 cm}
  \caption{Relative enhancement $T/T_0$ of the transmission as a function of
 the width $\lambda_w$ of the domain wall for various impurity concentrations
 $c$ and $L_x=8~\lambda_F$. The circles, squares and diamonds correspond to
$c = 3.85~\%$, $c = 5.77~\%$ and $c=7.69~\%$ respectively.
The solid line depicts the theoretical result for $c = 7.69~\%$
 ($\alpha = 0.02$, $\beta=6$).} \label{Walls2}
\end{center}
\end{figure}

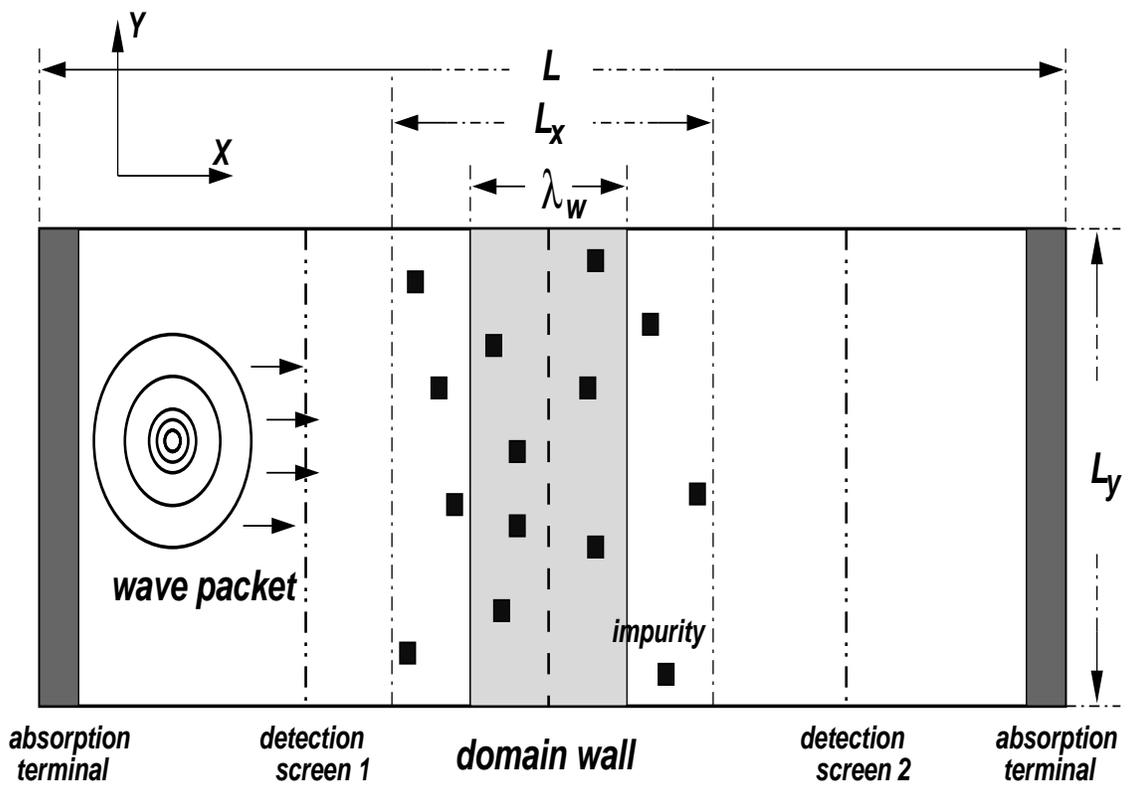
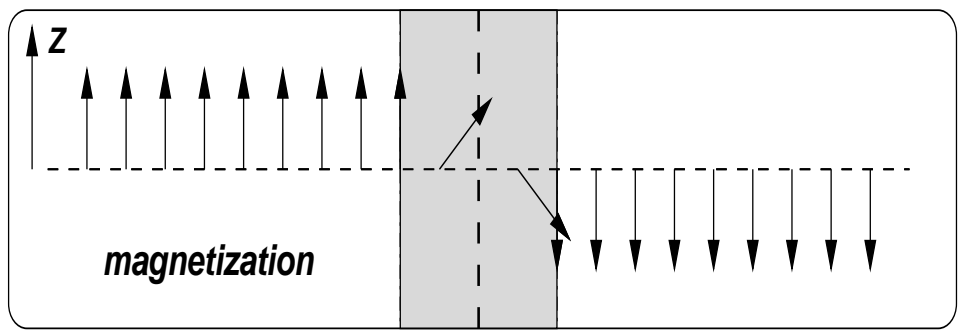
\end{document}